\documentclass{article}

\usepackage{iftex}
\ifPDFTeX
  \usepackage[T1]{fontenc}
  \usepackage[utf8]{inputenc}
  \usepackage{lmodern}
\fi
\usepackage[margin=1in]{geometry}
\usepackage{amsmath,amssymb,amsfonts}
\usepackage{booktabs}
\usepackage{caption}
\usepackage{subcaption}
\usepackage{natbib}
\usepackage{xurl}
\usepackage{siunitx}
\usepackage{algorithm}
\usepackage{algpseudocode}
\usepackage{float}
\usepackage{placeins}
\usepackage{etoolbox}
\ifPDFTeX
  \usepackage{graphicx}
  \usepackage{hyperref}
\else
  \ifXeTeX
    \usepackage{graphicx}
    \usepackage{hyperref}
  \else
    \ifLuaTeX
      \usepackage{graphicx}
      \usepackage{hyperref}
    \else
      \usepackage[dvipdfmx]{graphicx}
      \usepackage[dvipdfmx]{hyperref}
    \fi
  \fi
\fi

\AtBeginDocument{%
  \pretocmd{\section}{\FloatBarrier}{}{}%
  \pretocmd{\subsection}{\FloatBarrier}{}{}%
}
\hypersetup{
  colorlinks=true,
  linkcolor=blue,
  citecolor=blue,
  urlcolor=blue
}

\newcommand{\KL}{\mathrm{KL}}
\newcommand{\E}{\mathbb{E}}

\newcommand{\R}{\mathbb{R}}
\newcommand{\eps}{\varepsilon}
\newcommand{\HVA}{\mathrm{HVA}}
\newcommand{\CVaR}{\mathrm{CVaR}}
\newcommand{\ESS}{\mathrm{ESS}}
\newcommand{\TE}{\mathrm{TE}}
\newcommand{\ind}{\mathrm{ind}}

\title{Robust Hedging Valuation Adjustment under Liquidity--Demand Stress}
\author{Takayuki Sakuma\footnote{e-mail: tsakuma@soka.ac.jp.}\\
Soka University}
\date{}

\begin{document}
\maketitle

\begin{abstract}
This paper develops a robust hedging valuation adjustment (HVA) measure for dynamic hedging. Simulated rebalancing and maturity-unwind trades generate a loss distribution for each no-trade-band rule, and we define robust HVA as the worst-case expected loss over a relative-entropy neighborhood of that distribution. Because band width affects turnover, the same relative-entropy radius applied to different bands can imply different levels of demand--liquidity stress. We distinguish a fixed-radius convention from a fixed benchmark-stress convention and show that wider no-trade bands lower rebalancing costs but raise hedge-error risk.
\end{abstract}


\section{Introduction}\label{sec:intro}
Dynamic hedging is not frictionless: discrete rebalancing, bid--ask spreads, and market impact can make even standard derivative hedges costly. Burnett \citep{Burnett2020HVA} formulates hedging valuation adjustment (HVA) as the effect of transaction costs on derivative portfolio value. Burnett and Williams \citep{BurnettWilliams2021CostHedgingXVA} extend this idea to the cost of hedging XVA, including frictions from hedging the underlying asset and counterparty credit. Related HVA research also examines frictions beyond direct transaction costs. \citet{AlbaneseBenezetCrepey2022HVAModelRisk} link HVA to model risk, \citet{BenezetCrepey2024HandlingModelRiskXVAs} discuss its role in XVA model-risk management, and \citet{BenezetCrepeyEssaket2026CallableHVA} study callable-product settings where recalibration and hedging adjustments interact.

Our focus on liquidity costs relates to the literature on hedging with transaction costs. \citet{Leland1985} studies modified-volatility replication under transaction costs. \citet{DavisPanasZariphopoulou1993} develop utility-based valuation and hedging. \citet{WhalleyWilmott1997} derive small-transaction-cost asymptotic approximations for the no-transaction region. \citet{SonerShreveCvitanic1995} show that nontrivial replication breaks down under proportional transaction costs. Jump risk is also an issue for hedging with frictions. \citet{KennedyForsythVetzal2009} study dynamic hedging under jump diffusion with transaction costs and \citet{Sepp2012DeltaHedgingErrors} derives approximate delta-hedging error distributions under jump diffusion with discrete trading and transaction costs.

In this paper, we turn hedging costs into stressed valuation adjustments by applying entropy-based robust risk measurement. For risk measurement, Kullback--Leibler (KL), or relative-entropy, neighborhoods around a baseline distribution provide tractable worst-case expectations \citep{GlassermanXu2014}. We compare robust HVA across hedge policies under liquidity and demand stress, and our approach lets each policy generate its own turnover, hedge-error risk, and empirical loss distribution. We focus on no-trade-band delta hedging, where the hedge is updated only when the target delta moves beyond a specified band width. The costs of rebalancing define the empirical HVA loss sample for that policy. This policy dependence creates a practical comparison issue. Applying a different band to the same simulated scenarios alters turnover, the hedge-demand measure, and the empirical HVA loss sample. A common relative-entropy radius fixes the admissible probability tilt but not necessarily the economic demand and liquidity stress. A lower robust HVA for one band might merely reflect a lower effective stress level, rather than a genuinely safer hedge policy. We therefore introduce two measures. A fixed-radius convention uses the same relative-entropy radius for every hedge band. A fixed benchmark-stress convention instead fixes the demand--liquidity stress label and allows the required KL radius to vary. Our numerical results indicate that valuations under fixed-radius and benchmark-stress conventions are close in high-liquidity environments but diverge when liquidity is low. Wider no-trade bands reduce rebalancing costs and robust HVA, though they increase hedge-error risk.

\section{No-trade-band policy}\label{sec:hva_model}

Let $H(S_T)$ be the derivative payoff at maturity $T$ and let $S$ be the underlying asset. The numerical experiments use an at-the-money European call, $H(S_T)=(S_T-K)^+$ with $K=S_0$. The baseline hedge is the Black--Scholes--Merton delta \citep{BlackScholes1973,Merton1973},
\begin{equation}\label{eq:bs_delta}
  V(t,S)=S\Phi(d_1)-K e^{-r(T-t)}\Phi(d_2),
  \qquad
  \widehat \Delta_t=\partial_S V(t,S_t)=\Phi(d_1),
\end{equation}
where
\[
d_{1,2}=
\frac{\log(S/K)+(r\pm \tfrac12\sigma^2)(T-t)}
     {\sigma\sqrt{T-t}} .
\]
Consider rebalancing dates $0=t_0<t_1<\cdots<t_n=T$. The band width $b\ge 0$ is measured in units of underlying per option. Let $\Delta_{t_i}(b)$ be the hedge position held after rebalancing at $t_i$. Initializing $\Delta_{t_0}(b)=\widehat\Delta_{t_0}$ and for $i=1,\ldots,n-1$,
\begin{equation}\label{eq:band_policy}
\Delta_{t_i}(b)=
\begin{cases}
\widehat\Delta_{t_i}, & \text{if }|\widehat\Delta_{t_i}-\Delta_{t_{i-1}}(b)|>b,\\
\Delta_{t_{i-1}}(b), & \text{otherwise.}
\end{cases}
\end{equation}
At maturity the hedge is unwound by setting
\[
\Delta_{t_n}(b)=0.
\]
We define the rebalancing trade at time $t_i$ as the change in the underlying position,
\[
\Delta q_i(b)=\Delta_{t_i}(b)-\Delta_{t_{i-1}}(b).
\]
A positive $\Delta q_i(b)$ is a purchase of the underlying, and turnover is measured by the trading volume
\[
X_i(b)=S_{t_i}|\Delta q_i(b)|.
\]

\section{Kullback--Leibler robust HVA}\label{sec:hva_loss_sample}\label{sec:kl}

Let $m_{t_i}\ge 1$ denote the pathwise illiquidity multiplier at the rebalancing date $t_i$. The value $m_{t_i}=1$ indicates normal liquidity, while higher values indicate poorer liquidity. In the numerical liquidity environments, $m_{t_i}$ is treated as exogenous; Section~\ref{sec:design} outlines the two-state normal--stress specification.

Let $s$ represent the baseline half-spread and let $\kappa$ denote the quadratic impact coefficient. The discounted hedging-friction loss along a single path is given by
\begin{equation}\label{eq:hva_loss}
  L(b)
  =
  \sum_{i=1}^{n} e^{-r t_i}\,m_{t_i}\Big(sX_i(b)+\kappa X_i(b)^2\Big).
\end{equation}
Equation~\eqref{eq:hva_loss} defines the paper’s HVA loss functional. It implements the HVA idea in a discrete-time transaction-cost setting: rebalancing and unwind frictions enter the valuation adjustment as hedge costs \citep{Burnett2020HVA,BurnettWilliams2021CostHedgingXVA}. The term $sX_i(b)$ represents the linear spread cost and $\kappa X_i(b)^2$ reflects the convex temporary market impact. Rebalancing trades correspond to $i=1,\ldots,n-1$, and the maturity unwind corresponds to $i=n$. The position $\Delta_{t_0}(b)=\widehat\Delta_{t_0}$ is not charged.

Let $P^{\ind}$ be the baseline simulation law under which the liquidity multiplier process is generated independently of prices. The baseline mean HVA of policy $b$ under this law is the expected discounted loss value,
\begin{equation}\label{eq:hva_ind}
\HVA^{\ind}(b)=\E_{P^{\ind}}[L(b)]
\end{equation}
and $P^{\ind}$ is represented by an equally weighted Monte Carlo sample,
\[
L_1(b),\ldots,L_N(b).
\]
The robust HVA framework allows the scenario weights to vary within a relative-entropy neighborhood. For a radius \(\eps\ge 0\), we define
\begin{equation}\label{eq:kl_ball}
  \mathcal U_\eps
  =
  \left\{
  w\in\R_+^N:
  \sum_{j=1}^N w_j=1,\quad
  \sum_{j=1}^N w_j\log(Nw_j)\le \eps
  \right\},
\end{equation}
where \(w^0=(1/N,\ldots,1/N)\) denotes the baseline empirical weights.
The expression in \eqref{eq:kl_ball} represents the relative entropy of \(w\) with respect to \(w^0\), and the radius \(\eps\) determines the allowed probability tilt. A larger \(\eps\) allows higher weight to be placed on high-loss scenarios.

We define the KL-robust upper HVA for band \(b\) as
\begin{equation}\label{eq:hva_upper}
  \overline{\HVA}_\eps(b)
  =
  \sup_{w\in\mathcal U_\eps}\sum_{j=1}^N w_jL_j(b).
\end{equation}
The KL-robust expectation has the entropic dual representation \citep{GlassermanXu2014}:
\begin{equation}\label{eq:dual}
  \overline{\HVA}_\eps(b)
  =
  \inf_{\theta>0}
  \left\{
  \theta\eps+
  \theta\log\left(\frac{1}{N}\sum_{j=1}^N \exp(L_j(b)/\theta)\right)
  \right\}.
\end{equation}
For a fixed Lagrange multiplier \(\theta>0\) on the entropy constraint and \(\eta\) on the probability normalization constraint, the Lagrangian is
\[
  \sum_{j=1}^N w_jL_j(b)
  -\theta\sum_{j=1}^N w_j\log(Nw_j)
  +\eta\left(\sum_{j=1}^N w_j-1\right).
\]
Differentiating with respect to \(w_j\) gives
\[
  L_j(b)-\theta\{\log(Nw_j)+1\}+\eta=0.
\]
Thus \(w_j\) is proportional to \(\exp(L_j(b)/\theta)\), and normalizing and evaluating at \(\theta^\star\) gives
\begin{equation}\label{eq:weights}
  w_j^\star(b,\eps)
  =
  \frac{\exp(L_j(b)/\theta^\star)}
       {\sum_{k=1}^{N}\exp(L_k(b)/\theta^\star)}.
\end{equation}

\section{Robust HVA increment}\label{sec:hva_increment}\label{sec:calibration}

We use two stress measures. The first is the Kullback--Leibler radius \(\eps\), which controls how far the worst-case scenario weights may move from the baseline empirical weights. The second is the Gaussian benchmark-stress label \(\rho^G\), which translates the KL stress increment into a demand--illiquidity stress.

The Gaussian benchmark is a one-parameter translation convention. The parameter \(\rho^G\) changes the rank pairing between turnover and illiquidity while preserving their marginal distributions. Kendall's \(\tau\) from data is mapped to the Gaussian rank-correlation scale through \(\rho=\sin(\pi\tau/2)\). This distinction is important because the KL stress is applied to the loss sample generated by each hedge policy. Different policies lead to different trading paths, and thus to varying turnover, trade-size, and distributions. Consequently, applying the same numerical KL radius across policies can mix the chosen stress severity with policy-driven changes in the loss distribution.

The KL stress increment over the baseline HVA for band \(b\) is
\begin{equation}\label{eq:hva_increment}
  \Delta_b^{\KL}(\eps)
  =
  \overline{\HVA}_\eps(b)-\overline{\HVA}_0(b),\overline{\HVA}_0(b)=\HVA^\ind(b).
\end{equation}
To interpret this increment, we compare it with a Gaussian rank-coupling benchmark. For each band, the benchmark keeps the simulated marginal distributions of turnover and illiquidity fixed, but changes their rank pairing. When \(\rho^G=0\), high-turnover paths are not consistently paired with high-illiquidity paths. As \(\rho^G\) increases, high-turnover paths become more frequently paired with high-illiquidity paths, increasing the
benchmark HVA cost.

Let \(C_b^G(\rho^G)\) represent the average HVA cost in this Gaussian benchmark; Appendix~\ref{app:gaussian_benchmark} describes how \(C_b^G(\rho^G)\) is constructed. We define the benchmark increment as
\begin{equation}\label{eq:gaussian_increment_main}
  \Delta_b^G(\rho^G)=C_b^G(\rho^G)-C_b^G(0).
\end{equation}
Hence \(\Delta_b^G(\rho^G)\) measures the increase in benchmark HVA cost when the turnover and illiquidity rank coupling changes from \(0\) to \(\rho^G\).

The finite-sample benchmark increment curve consists of the grid \(\rho^G_j\mapsto\Delta_b^G(\rho^G_j)\), and Monte Carlo noise can cause this grid to be nonmonotone. The inverse map therefore applies the monotone envelope
\[
  \widetilde\Delta_b^G(\rho^G_j)=\max_{k\le j}\Delta_b^G(\rho^G_k)
\]
and the benchmark-stress label linked to a KL increment is
\begin{equation}\label{eq:stress_equiv}
  \rho^G_{\mathrm{eq}}(b;\eps)
  =
  \inf\{\rho^G:\widetilde\Delta_b^{G}(\rho^G)
       \ge \Delta_b^{\KL}(\eps)\}.
\end{equation}
Conversely, the KL radius required to match a benchmark-stress level \(\rho^G_0\) is
\begin{equation}\label{eq:eps_req}
  \eps_{\mathrm{req}}(b;\rho^G_0)
  =
  \inf\{\eps:\Delta_b^{\KL}(\eps)
       \ge \widetilde\Delta_b^{G}(\rho^G_0)\}.
\end{equation}
These definitions provide two conventions for comparing policies. The fixed-radius convention sets \(\eps(b)=\eps_{\mathrm{fixed}}\) for all \(b\). The fixed benchmark-stress convention sets \(\eps(b)=\eps_{\mathrm{req}}(b;\rho^G_0)\). The first keeps the admissible probability tilt constant. The second keeps the benchmark-stress level constant and allows the required KL radius to vary by band. It does not keep a common dollar increment constant across bands, since each band has its own turnover and effective-liquidity marginal scales.

In practice, a risk committee or model-governance process can set the stress label \(\rho^G_0\). This determines the severity of the demand--illiquidity stress used to compare hedge policies. For each band \(b\), the benchmark increment \(\widetilde{\Delta}_b^G(\rho^G_0)\) is calculated from that band's turnover and effective-illiquidity marginal scales. The required KL radius \(\eps_{\mathrm{req}}(b;\rho^G_0)\) is then selected so that the KL stress increment equals this benchmark increment. Hence \(\rho^G_0\) is common across policies, while \(\eps_{\mathrm{req}}\), robust HVA, and hedge-error risk remain specific to each policy. The resulting robust HVA represents the reserve cost of a hedge policy under a common benchmark-stress label, and it should be considered together with the tracking-error measure \(\TE(b)\) defined in Section~\ref{sec:band_selection}.

\section{Numerical results}\label{sec:results}

\subsection{Setting}\label{sec:design}
Table~\ref{tab:baseline_spec} gives the common settings. The band grid ranges from daily rebalancing ($b=0$) to a wide no-trade band ($b=0.5$). The three liquidity environments are generated with a risk-neutral Merton jump-diffusion \citep{Merton1976},
\begin{equation}\label{eq:merton}
  \frac{dS_t}{S_{t-}}
  =
  (r-\lambda_J\kappa_J)\,dt+\sigma_m  a_m\,dW_t+(J-1)\,dN_t,
  \qquad
  \log J\sim N(\mu_J,\sigma_J^2),
\end{equation}
where $N_t$ has intensity $\lambda_J$ and $\kappa_J=\exp(\mu_J+\tfrac12\sigma_J^2)-1$. Table~\ref{tab:enh_markets} gives the liquidity environments. They vary the half-spread, the quadratic impact coefficient, the stress multiplier, liquidity persistence, diffusion volatility, and the jump parameters. The high-liquidity case assumes a small half-spread with almost no impact. The medium- and low-liquidity cases use larger $s$ and $\kappa$ to represent more expensive trading. The high-liquidity environment sets $\lambda_J=0$ and therefore reduces to a diffusion benchmark. In an implementation, these inputs could be replaced by desk- or asset-class-specific estimates of spreads and temporary market impact. In each environment, the liquidity multiplier follows a two-state normal--stress process with values $1$ and $m_{\mathrm{stress}}$. The transition probabilities $p_{NN}$ and $p_{SS}$ in Table~\ref{tab:enh_markets} control persistence in the normal and stress states.

\begin{table}[H]
\centering
\caption{Baseline HVA specification.}
\label{tab:baseline_spec}
\begin{tabular}{p{0.46\linewidth} p{0.48\linewidth}}
\toprule
Item & Value \\
\midrule
Instrument & European call option \\
Spot $S_0$ & 1.0 \\
Strike $K$ & 1.0 \\
Maturity $T$ & 1.0 year \\
Risk-free rate $r$ & 0.02 \\
Volatility $\sigma_m$  & 0.2 \\
Time steps ($n_\mathrm{steps}$) & 252 (dt=0.0040) \\
Monte Carlo paths & 20,000 \\
Band grid & 0.0, 0.005, 0.01, 0.02, 0.05, 0.1, 0.2, 0.3, 0.5 \\
KL radius grid & 0.0, 0.001, 0.002, 0.005, 0.01, 0.02, 0.05, 0.1 \\
\bottomrule
\end{tabular}

\end{table}

\begin{table}[H]
\centering
\caption{Liquidity environments.}
\label{tab:enh_markets}
\resizebox{\textwidth}{!}{%
\begin{tabular}{lrrrrrrrrr}
\toprule
Liquidity environment &  $s$ &  $\kappa$ & $m_\mathrm{stress}$ & $p_{NN}$ & $p_{SS}$ & $a_m$  &  $\lambda_J$ &  $\mu_J$ &  $\sigma_J$ \\
\midrule
High-liquidity & 0.0001 & 0 & 5 & 0.985 & 0.75 & 1 & 0 & -0.05 & 0.1 \\
Medium-liquidity & 0.0005 & 0.0015 & 10 & 0.975 & 0.9 & 1.25 & 0.5 & -0.06 & 0.12 \\
Low-liquidity & 0.0025 & 0.008 & 20 & 0.96 & 0.95 & 1.5 & 2 & -0.1 & 0.18 \\
\bottomrule
\end{tabular}%
}

\end{table}

\subsection{Reference benchmark-stress level}\label{sec:public_calibration}
The fixed benchmark-stress calculations use \(\rho^G_0=0.4\). In the Gaussian benchmark, \(\rho^G\) represents how closely demand and illiquidity ranks move together. In the simulated HVA samples, demand arises from hedge turnover. Since trade-level hedge turnover is not available in public daily data, we rely on a related public-data reference: the co-movement between market stress and market illiquidity.

We use the CBOE VIX series from FRED as the measure of market stress \citep{FREDVIX}. Daily SPY prices and volumes are taken from Stooq \citep{StooqSPY}. Using the SPY data, we compute the Amihud illiquidity measure
\[
  \mathrm{ILLIQ}_t=\frac{|r_t|}{P_t\times \mathrm{volume}_t},
\]
following \citet{Amihud2002}, where \(r_t=P_t/P_{t-1}-1\) denotes the daily SPY return. The VIX and SPY Amihud series are matched by trading day.

The calculation uses 756-trading-day rolling windows, roughly three years, and shifts the window forward every 21 trading days. The first window spans from 2005-02-28 to 2008-02-28, and the last one extends from 2023-01-05 to 2026-01-09. For each window, we calculate Kendall's rank correlation \(\tau\) between VIX and the SPY Amihud measure. Since the Gaussian benchmark in Section~\ref{sec:calibration} is defined by the Gaussian rank-correlation \(\rho^G\), Kendall's \(\tau\) is translated to the Gaussian scale as
\[
  \rho^{\mathrm{pub}}=\sin(\pi\tau/2).
\]
The relevant stress case involves positive co-movement between market stress and illiquidity, so we keep
\[
  \rho^{\mathrm{pub}}_{\mathrm{wwr}}=\max(\rho^{\mathrm{pub}},0).
\]
Next, we focus on high-stress windows, defined as rolling windows whose average VIX falls within the top decile of all rolling-window averages. The public-data reference value corresponds to the 90th percentile of \(\rho^{\mathrm{pub}}_{\mathrm{wwr}}\) within those windows. In the 756-trading-day calculation, this yields \(\rho_{\mathrm{target}}\approx0.364\); the corresponding 95th percentile is approximately 0.368. We use \(\rho^G_0=0.4\) as a rounded benchmark stress level. Shorter rolling windows produce higher reference values, and Appendix~\ref{app:calib_audit} shows the sensitivity.

\begin{figure}[H]
\centering
\includegraphics[width=0.86\linewidth]{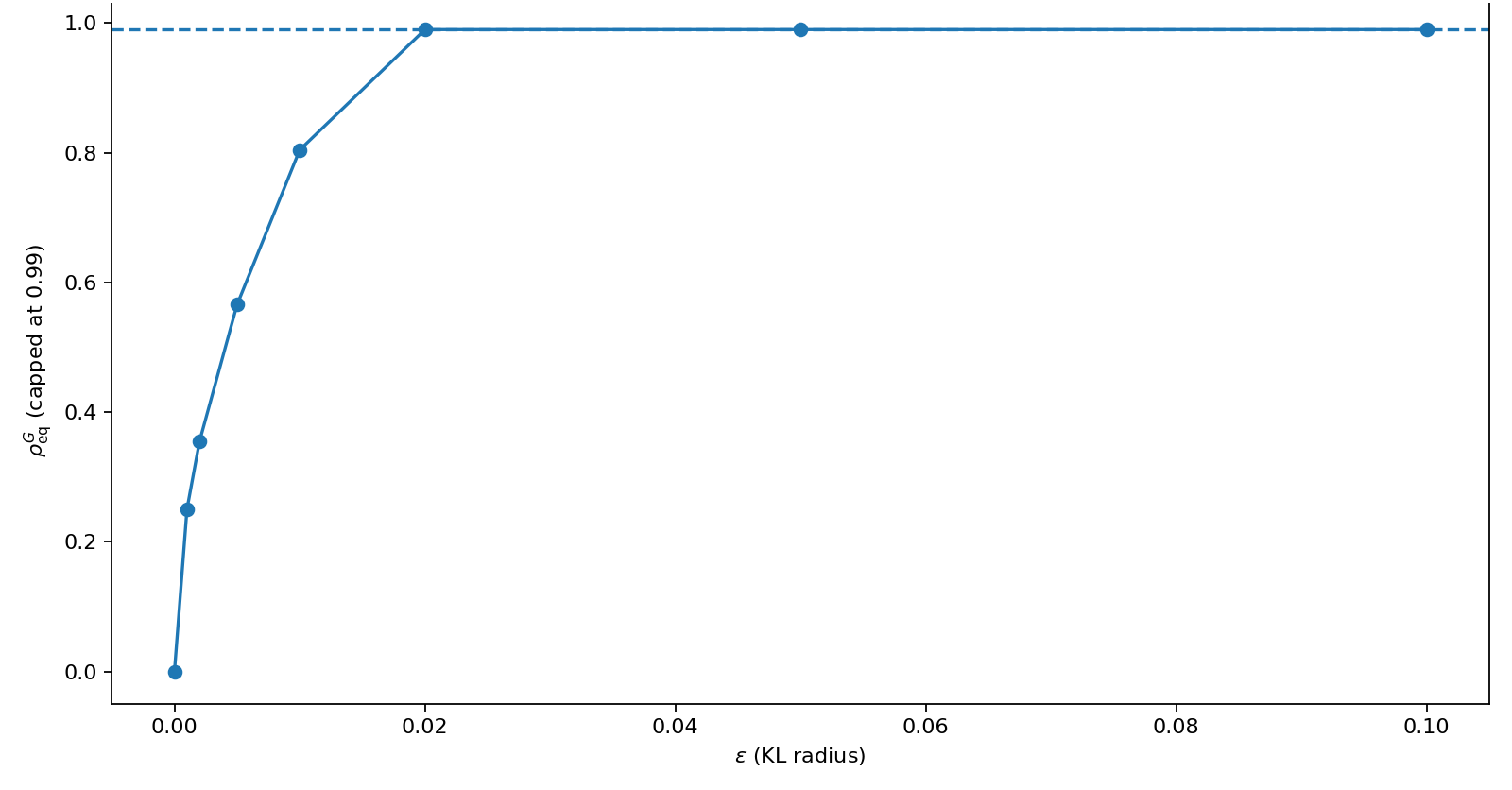}
\caption{Public-data map from Kullback--Leibler radius \(\eps\) to equivalent benchmark-stress label \(\rho^G_{\mathrm{eq}}(\eps)\). Values at the top grid boundary indicate boundary-limited matches.}
\label{fig:epsmap}
\end{figure}

Figure~\ref{fig:epsmap} shows the Kullback--Leibler radius \(\eps\) and the equivalent Gaussian benchmark-stress label \(\rho^G_{\mathrm{eq}}(\eps)\). For each \(\eps\), the KL stress increment \(\Delta^{\KL}(\eps)\) is aligned with the monotone Gaussian benchmark increment curve so that
\[
  \widetilde\Delta^G\big(\rho^G_{\mathrm{eq}}(\eps)\big)
  =
  \Delta^{\KL}(\eps).
\]
When \(\eps\) becomes large, the matched value reaches the top boundary of the benchmark grid; this is therefore reported as a boundary value, not as an exact structural value. Appendix~\ref{app:stress_level_sensitivity} discusses whether the main policy conclusions depend on the specific choice \(\rho^G_0=0.4\).

\subsection{Fixed-radius versus fixed benchmark-stress}\label{sec:fixed_views}
Table~\ref{tab:epsradii} reports the fixed radii used in the fixed-radius comparison. For each liquidity environment, the radius is calibrated at $b=0$ by solving $\Delta_0^{\KL}(\eps)=\widetilde\Delta_0^G(\rho^G_0)$ with $\rho^G_0=0.4$.

\begin{table}[H]
\centering
\caption{Fixed KL radii.}
\label{tab:epsradii}
\begin{tabular}{lr}
\toprule
Liquidity environment & $\varepsilon_{\mathrm{fixed}}$ \\
\midrule
High-liquidity & 0.0029 \\
Medium-liquidity & 0.0036 \\
Low-liquidity & 0.0038 \\
\bottomrule
\end{tabular}

\end{table}

Using the fixed radii in Table~\ref{tab:epsradii}, Table~\ref{tab:fixedradius_label_decay} reports the matched $\rho^G_{\mathrm{eq}}(b;\eps_{\mathrm{fixed}})$ for selected bands. The fixed benchmark-stress calculation solves for $\eps_{\mathrm{req}}(b;\rho^G_0)$, the radius needed to maintain $\rho^G_0=0.4$ for that band. At $b=0$, the two results coincide but away from $b=0$, they diverge. The divergence is small in the high-liquidity environment but grows as liquidity deteriorates. 

\begin{table}[H]
\centering
\caption{Fixed-radius interpretation at selected bands.}
\label{tab:fixedradius_label_decay}
\resizebox{\textwidth}{!}{%
\begin{tabular}{lrrrrr}
\toprule
Liquidity environment & $b$ & $\rho^G_{\mathrm{eq}}(b;\varepsilon_{\mathrm{fixed}})$ & $\varepsilon_{\mathrm{req}}/\varepsilon_{\mathrm{fixed}}$ & $\overline{\HVA}_{\varepsilon_{\mathrm{fixed}}}(b)$ & $\overline{\HVA}_{\varepsilon_{\mathrm{req}}}(b)$ \\
\midrule
High-liquidity & 0 & 0.400 & 1.00 & \num{0.00074} & \num{0.00074} \\
High-liquidity & 0.02 & 0.357 & 1.22 & \num{0.00063} & \num{0.00063} \\
High-liquidity & 0.5 & 0.317 & 1.54 & \num{8.71e-05} & \num{8.85e-05} \\
Medium-liquidity & 0 & 0.399 & 1.00 & \num{0.0132} & \num{0.0132} \\
Medium-liquidity & 0.1 & 0.334 & 1.39 & \num{0.0081} & \num{0.0082} \\
Medium-liquidity & 0.5 & 0.263 & 2.29 & \num{0.0029} & \num{0.0031} \\
Low-liquidity & 0 & 0.400 & 1.00 & \num{0.244} & \num{0.244} \\
Low-liquidity & 0.3 & 0.208 & 3.03 & \num{0.1287} & \num{0.1396} \\
Low-liquidity & 0.5 & 0.252 & 2.43 & \num{0.0593} & \num{0.063} \\
\bottomrule
\end{tabular}%
}

\end{table}

\begin{figure}[H]
\centering
\includegraphics[width=0.86\linewidth]{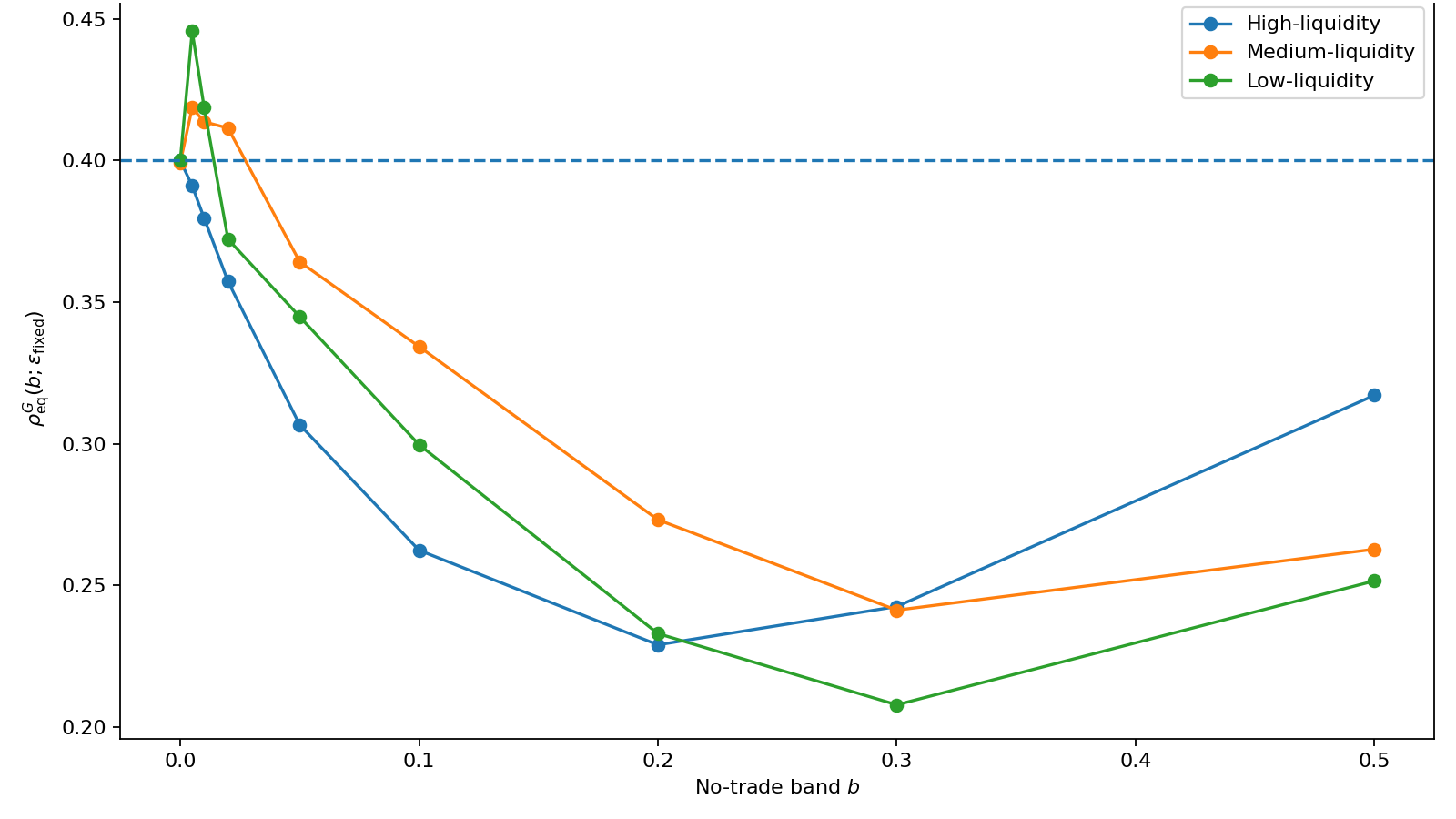}
\caption{Equivalent benchmark-stress label \(\rho^G_{\mathrm{eq}}(b;\eps_{\mathrm{fixed}})\) under the fixed-radius view.}
\label{fig:stress_label_band}
\end{figure}

\begin{figure}[H]
\centering
\includegraphics[width=0.86\linewidth]{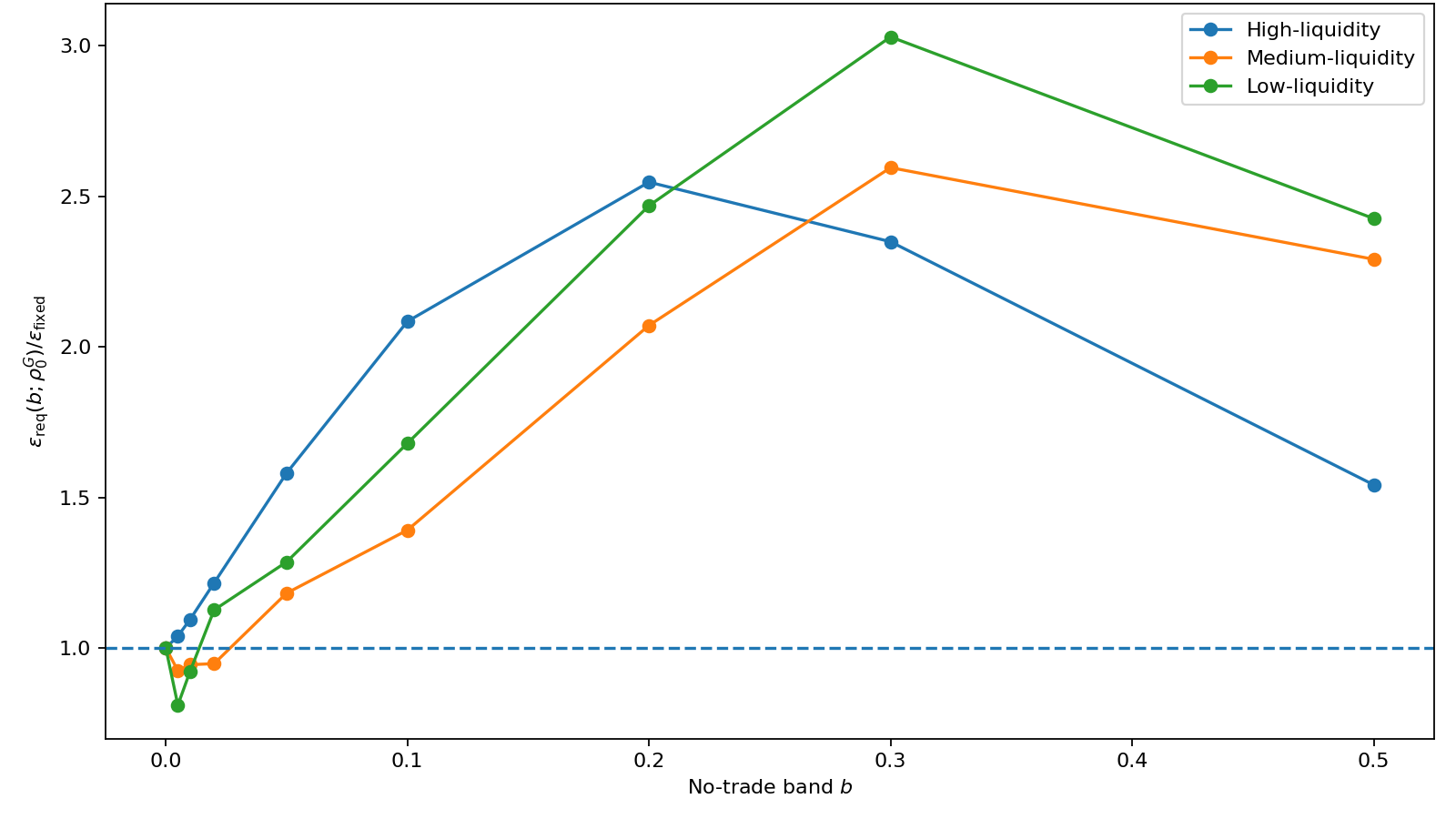}
\caption{Required KL-radius ratio $\eps_{\mathrm{req}}(b;\rho^G_0)/\eps_{\mathrm{fixed}}$ needed to keep $\rho^G_0=0.4$.}
\label{fig:eps_ratio_band}
\end{figure}

Figures~\ref{fig:stress_label_band} and \ref{fig:eps_ratio_band} apply the same calculation to the full band grid. Figure~\ref{fig:stress_label_band} plots the label implied by the fixed-radius calculation. Figure~\ref{fig:eps_ratio_band} plots the ratio $\eps_{\mathrm{req}}(b;\rho^G_0)/\eps_{\mathrm{fixed}}$ needed to keep the benchmark-stress label fixed. The fixed-radius interpretation is not invariant to policy-induced turnover; it may weaken when the band significantly changes trading demand. The required KL-radius ratio reaches its highest level in the low-liquidity case.

\subsection{Grid-selected no-trade bands}\label{sec:band_selection}
To isolate hedge error, we recompute the terminal replication error after setting bid--ask and market-impact costs to zero. \(\TE(b)\) reflects replication error only while HVA captures trading-friction costs. Let \(\tilde S_{t_i}=e^{-r t_i}S_{t_i}\) and let \(P_0^{\mathrm{ref}}\) denote the initial derivative value used in this computation. The discounted terminal value of the self-financing hedge induced by \(b\), before subtracting any transaction costs, is given by
\begin{equation}\label{eq:tracking_error}
\tilde V_T^{(0)}(b)
=
P_0^{\mathrm{ref}}
+
\sum_{i=0}^{n-1}\Delta_{t_i}(b)\big(\tilde S_{t_{i+1}}-\tilde S_{t_i}\big),
\qquad
E^{(0)}(b)=\tilde V_T^{(0)}(b)-e^{-rT}H(S_T),
\end{equation}
where \(P_0^{\mathrm{ref}}=V(0,S_0)\) in the high-liquidity environment and the Monte Carlo risk-neutral value \(P_0^{\mathrm{ref}}=N^{-1}\sum_j e^{-rT}H(S_{T,j})\) in the jump-diffusion stress environments. We define tracking-error risk as the 95\% CVaR of the replication shortfall:
\begin{equation}\label{eq:te_def}
  \TE(b)=\CVaR_{95\%}\big(-E^{(0)}(b)\big).
\end{equation}
To keep the tracking-error penalty comparable across liquidity regimes, the experiments set
\begin{equation}\label{eq:lambda_norm}
  \lambda=\lambda_\star\lambda_{\mathrm{norm}},
  \qquad
  \lambda_\star=\frac{\overline{\HVA}_0(0)}{\TE(0)}.
\end{equation}
Thus \(\lambda_{\mathrm{norm}}=1\) gives the \(b=0\) mean HVA and the \(b=0\) tracking-error risk equal weight at the reference point. Unless otherwise stated, the policy-selection tables use \(\lambda_{\mathrm{norm}}=1\). Alternative tracking-error weights appear in Appendix~\ref{app:lambda_sensitivity}.

For a policy view \(v\), let \(\HVA_v(b)\) denote the HVA measure used in that view. In the fixed benchmark-stress view, \(\HVA_v(b)=\overline{\HVA}_{\varepsilon_{\mathrm{req}}(b;\rho^G_0)}(b)\). The cost–risk trade-off is summarized by
\[
  J_v(b)=\HVA_v(b)+\lambda\TE(b),
  \qquad
  b_v^\star=\arg\min_b J_v(b).
\]
For each environment and band, we compute \(\HVA_v(b)\) and \(\TE(b)\), form \(J_v(b)\), and select the grid band \(b_v^\star\) with the lowest objective value. Table~\ref{tab:optband_fixedbenchmark} identifies \(b=0.02\) in the high-liquidity environment, \(b=0.10\) in the medium-liquidity environment, and the wide-band point \(b=0.30\) in the low-liquidity environment. The corresponding relative robust HVA increments are 3.77\%, 12.76\%, and 21.58\%. The low-liquidity objective surface appears almost flat across nearby wide bands. Further sensitivity to the tracking-error weight is discussed in Appendix~\ref{app:lambda_sensitivity}.

\begin{table}[H]
\centering
\caption{Grid-selected no-trade band under \(\rho^G_0=0.4\) and \(\lambda_{\mathrm{norm}}=1\).}
\label{tab:optband_fixedbenchmark}
\resizebox{\textwidth}{!}{%
\begin{tabular}{lrrrrrr}
\toprule
Liquidity environment & $b^\star$ & $\overline{\HVA}_0(b^\star)$ & $\overline{\HVA}_{\varepsilon_{\mathrm{req}}}(b^\star)$ & $\Delta_b^{\KL}$ & $\Delta_b^{\KL}/\overline{\HVA}_0$ (\%) & $\TE(b^\star)$ \\
\midrule
High-liquidity & 0.02 & \num{0.00061} & \num{0.00063} & \num{2.283e-05} & 3.77 & \num{0.0085} \\
Medium-liquidity & 0.1 & \num{0.00731} & \num{0.00824} & \num{0.00093} & 12.76 & \num{0.0654} \\
Low-liquidity & 0.3 & \num{0.1148} & \num{0.1396} & \num{0.02478} & 21.58 & \num{0.1547} \\
\bottomrule
\end{tabular}%
}

\end{table}

\subsection{Gamma sensitivity}\label{sec:gamma_rebalancing}
In this section, we let the band depend on gamma and discuss whether high gamma changes turnover, transaction costs, and robust HVA. We use delta hedging only; gamma is not hedged with an additional option or other convex instrument. High gamma makes the target delta more sensitive to underlying-price changes and therefore raises rebalancing demand. Letting $\Gamma_i$ be the Black--Scholes gamma, the first-order approximation gives
\[
  \Delta_{t+\Delta t}-\Delta_t\approx \Gamma_t(S_{t+\Delta t}-S_t).
\]
Combining this with the diffusion-scale standard deviation
\[
  \mathrm{sd}(S_{t+\Delta t}-S_t)\approx S_t\sigma_m a_m \sqrt{\Delta t}
\]
gives the one-step gamma measure
\begin{equation}\label{eq:gamma_intensity}
  G_i=|\Gamma_i|S_i\sigma_m a_m \sqrt{\Delta t}.
\end{equation}
Let $\widetilde G_i$ denote $G_i$ normalized by its 90th percentile. For a base band $b_0$, we compare the fixed band $b_i=b_0$ with two state-dependent variants,
\begin{align}
  b_i^{\mathrm{tight}}&=\max\{0.25b_0,\; b_0/(1+\widetilde G_i)\},\label{eq:gamma_tight}\\
  b_i^{\mathrm{wide}}&=\min\{3b_0,\; b_0(1+\widetilde G_i)\}.\label{eq:gamma_wide}
\end{align}
The tightened rule rebalances more actively in high-\(G_i\) observations. In Table~\ref{tab:gamma_policy_selected}, Turnover denotes the path-average total trading volume and Trades denotes the path-average number of rebalancing trades across the one-year hedge horizon. The gamma-widened rule lowers robust HVA and turnover in high- and medium-liquidity environments, yet it increases \(\TE(b)\). Conversely, the gamma-tightened rule decreases \(\TE(b)\) in medium- and low-liquidity environments but raises robust HVA and turnover.

\begin{table}[H]
\centering
\caption{Gamma no-trade-band sensitivity under the fixed benchmark-stress view, $\rho^G_0=0.4$ and $\lambda_{\mathrm{norm}}=1$.}
\label{tab:gamma_policy_selected}
\resizebox{\textwidth}{!}{\begin{tabular}{llllllll}
\toprule
Liquidity environment & Policy & Base band $b_0$ & Robust HVA & $\TE$ & Turnover & Trades & Top-10\% $G_i$ turnover share \\
\midrule
High-liquidity & Gamma-widened & 0.019 & 0.00050 & 0.01151 & 4.006 & 66.5  & 23.2\% \\
High-liquidity & Fixed band & 0.020 & 0.00059 & 0.01054 & 4.707 & 101.1  & 23.3\% \\
High-liquidity & Gamma-tightened & 0.023 & 0.00063 & 0.0105 & 5.044 & 129.4  & 22.7\% \\
\midrule
Medium-liquidity & Gamma-widened & 0.070 & 0.0076 & 0.05945 & 2.102 & 12.7  & 20.6\% \\
Medium-liquidity & Fixed band & 0.100 & 0.0078 & 0.05916 & 2.291 & 14.6  & 26.5\% \\
Medium-liquidity & Gamma-tightened & 0.100 & 0.0089 & 0.05692 & 3.112 & 33.1 & 29.6\% \\
\midrule
Low-liquidity & Fixed band & 0.250 & 0.1322 & 0.1593 & 1.462 & 4.0  & 18.3\% \\
Low-liquidity & Gamma-tightened & 0.210 & 0.1424 & 0.1523 & 2.102 & 11.7 & 29.5\% \\
Low-liquidity & Gamma-widened & 0.150 & 0.1317 & 0.1619 & 1.457 & 4.3  & 13.9\% \\
\bottomrule
\end{tabular}
}
\end{table}

\begin{figure}[H]
\centering
\includegraphics[width=0.82\linewidth]{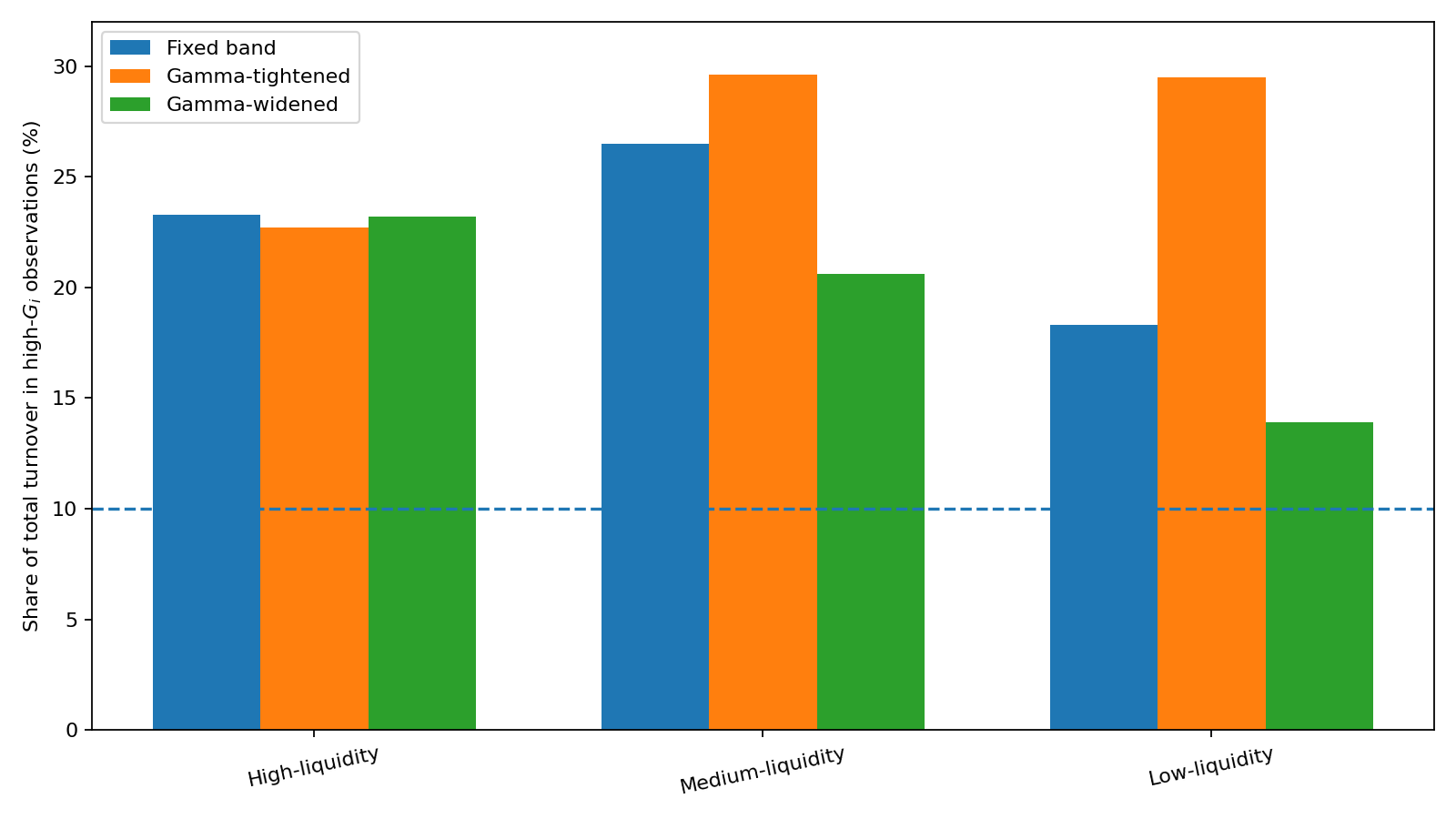}
\caption{Share of total turnover occurring when \(G_i\) is in the top decile of its empirical distribution. Values above 10\% indicate concentration of turnover in these high-\(G_i\) observations.}
\label{fig:gamma_turnover_share}
\end{figure}

Figure~\ref{fig:gamma_turnover_share} reports the share of total turnover that occurs when \(G_i\) is in the top decile of its empirical distribution. These cases account for 10\% of path-date observations. A share greater than 10\% indicates that high-\(G_i\) observations produce more turnover than their frequency by itself would suggest. Further weight-based measures for the KL tilt appear in Appendix~\ref{app:tail_measures}.

\section{Conclusion}\label{sec:conclusion}
This paper studies robust hedging valuation adjustment under liquidity--demand stress. The robust layer applies exponential tilting to simulated HVA loss samples generated by no-trade-band hedging policies. Because the hedge policy changes turnover, it also changes the empirical loss distribution to which the KL ball is applied. We distinguish two comparison conventions: a fixed-radius convention that keeps the relative-entropy radius constant across hedge bands, and a fixed benchmark-stress convention that keeps the benchmark demand--liquidity stress label constant while allowing the required KL radius to vary by band. The numerical results show that the difference is small in high-liquidity markets but larger under low liquidity. Wider no-trade bands lower rebalancing costs and robust HVA, yet they raise hedge-error risk. The gamma sensitivity analysis indicates that high-gamma delta movement accounts for a disproportionate share of rebalancing demand, so state-dependent bands can shift the balance between robust HVA and hedge-error risk. Future work could apply the same policy-comparison logic to other HVA settings, such as XVA hedging costs, model-risk HVA, and callable products, path-dependent derivatives, or other structured instruments.

\section*{Declaration of AI-assisted technologies}
During the preparation of this manuscript, the author used generative AI and AI-assisted tools to assist with manuscript drafting, language refinement, code drafting and debugging, and discussion of the presentation of the research. The author directed the project, reviewed, revised, and verified the manuscript and code, and takes full responsibility for the final manuscript.

\bibliographystyle{plainnat}
\bibliography{references_hva}

@article{BlackScholes1973,
  author  = {Black, Fischer and Scholes, Myron},
  title   = {The Pricing of Options and Corporate Liabilities},
  journal = {Journal of Political Economy},
  year    = {1973},
  volume  = {81},
  number  = {3},
  pages   = {637--654},
  doi     = {10.1086/260062}
}

@article{Merton1973,
  author  = {Merton, Robert C.},
  title   = {Theory of Rational Option Pricing},
  journal = {The Bell Journal of Economics and Management Science},
  year    = {1973},
  volume  = {4},
  number  = {1},
  pages   = {141--183},
  doi     = {10.2307/3003143}
}

@article{Merton1976,
  author  = {Merton, Robert C.},
  title   = {Option Pricing When Underlying Stock Returns Are Discontinuous},
  journal = {Journal of Financial Economics},
  year    = {1976},
  volume  = {3},
  number  = {1--2},
  pages   = {125--144},
  doi     = {10.1016/0304-405X(76)90015-2}
}

@article{GlassermanXu2014,
  author  = {Glasserman, Paul and Xu, Xingbo},
  title   = {Robust Risk Measurement and Model Risk},
  journal = {Quantitative Finance},
  year    = {2014},
  volume  = {14},
  number  = {1},
  pages   = {29--58},
  doi     = {10.1080/14697688.2013.822989}
}

@techreport{Burnett2020HVA,
  author      = {Burnett, Benedict},
  title       = {Hedging Valuation Adjustment: Fact and Friction},
  institution = {SSRN},
  year        = {2020},
  month       = oct,
  note        = {Available at SSRN: \url{https://ssrn.com/abstract=3709477}},
  doi         = {10.2139/ssrn.3709477}
}

@techreport{BurnettWilliams2021CostHedgingXVA,
  author      = {Burnett, Benedict and Williams, Ieuan},
  title       = {The Cost of Hedging XVA},
  institution = {SSRN},
  year        = {2021},
  month       = feb,
  note        = {Available at SSRN: \url{https://ssrn.com/abstract=3777095}},
  doi         = {10.2139/ssrn.3777095}
}

@techreport{AlbaneseBenezetCrepey2022HVAModelRisk,
  author      = {Albanese, Claudio and B{\'e}n{\'e}zet, Cyril and Cr{\'e}pey, St{\'e}phane},
  title       = {Hedging Valuation Adjustment and Model Risk},
  institution = {Working paper},
  year        = {2022},
  month       = nov,
  note        = {Version dated November 18, 2022. Available at \url{https://perso.lpsm.paris/~crepey/papers/HVA-model.pdf}}
}

@article{BenezetCrepey2024HandlingModelRiskXVAs,
  author  = {B{\'e}n{\'e}zet, Cyril and Cr{\'e}pey, St{\'e}phane},
  title   = {Handling Model Risk with XVAs},
  journal = {Frontiers of Mathematical Finance},
  year    = {2024},
  volume  = {3},
  number  = {4},
  pages   = {490--519},
  doi     = {10.3934/fmf.2024016}
}

@article{Amihud2002,
  author  = {Amihud, Yakov},
  title   = {Illiquidity and Stock Returns: Cross-Section and Time-Series Effects},
  journal = {Journal of Financial Markets},
  year    = {2002},
  volume  = {5},
  number  = {1},
  pages   = {31--56},
  doi     = {10.1016/S1386-4181(01)00024-6}
}

@misc{FREDVIX,
  author       = {{Chicago Board Options Exchange}},
  title        = {{CBOE} Volatility Index: {VIX} [{VIXCLS}]},
  howpublished = {Federal Reserve Bank of St. Louis, FRED},
  year         = {2026},
  note         = {Retrieved March 24, 2026, from \url{https://fred.stlouisfed.org/series/VIXCLS}}
}

@misc{StooqSPY,
  author       = {{Stooq.com}},
  title        = {{SPY.US} Historical Daily {OHLCV} Data},
  howpublished = {\url{https://stooq.com}},
  year         = {2026},
  note         = {Accessed March 25, 2026}
}

@misc{BenezetCrepeyEssaket2026CallableHVA,
  author       = {B{\'e}n{\'e}zet, Cyril and Cr{\'e}pey, St{\'e}phane and Essaket, Dounia},
  title        = {The Recalibration Conundrum: {Hedging Valuation Adjustment} for Callable Claims},
  year         = {2026},
  note         = {arXiv:2304.02479},
  doi          = {10.48550/arXiv.2304.02479}
}

@article{Leland1985,
  author  = {Leland, Hayne E.},
  title   = {Option Pricing and Replication with Transactions Costs},
  journal = {The Journal of Finance},
  year    = {1985},
  volume  = {40},
  number  = {5},
  pages   = {1283--1301},
  doi     = {10.1111/j.1540-6261.1985.tb02383.x}
}

@article{DavisPanasZariphopoulou1993,
  author  = {Davis, Mark H. A. and Panas, Vassilios G. and Zariphopoulou, Thaleia},
  title   = {European Option Pricing with Transaction Costs},
  journal = {SIAM Journal on Control and Optimization},
  year    = {1993},
  volume  = {31},
  number  = {2},
  pages   = {470--493},
  doi     = {10.1137/0331022}
}

@article{SonerShreveCvitanic1995,
  author  = {Soner, H. Mete and Shreve, Steven E. and Cvitani{\'c}, Jak{\v{s}}a},
  title   = {There Is No Nontrivial Hedging Portfolio for Option Pricing with Transaction Costs},
  journal = {The Annals of Applied Probability},
  year    = {1995},
  volume  = {5},
  number  = {2},
  pages   = {327--355}
}

@article{WhalleyWilmott1997,
  author  = {Whalley, A. Elizabeth and Wilmott, Paul},
  title   = {An Asymptotic Analysis of an Optimal Hedging Model for Option Pricing with Transaction Costs},
  journal = {Mathematical Finance},
  year    = {1997},
  volume  = {7},
  number  = {3},
  pages   = {307--324},
  doi     = {10.1111/1467-9965.00034}
}

@article{KennedyForsythVetzal2009,
  author  = {Kennedy, J. S. and Forsyth, P. A. and Vetzal, K. R.},
  title   = {Dynamic Hedging Under Jump Diffusion with Transaction Costs},
  journal = {Operations Research},
  year    = {2009},
  volume  = {57},
  number  = {3},
  pages   = {541--559},
  doi     = {10.1287/opre.1080.0598}
}

@article{Sepp2012DeltaHedgingErrors,
  author  = {Sepp, Artur},
  title   = {An Approximate Distribution of Delta-Hedging Errors in a Jump-Diffusion Model with Discrete Trading and Transaction Costs},
  journal = {Quantitative Finance},
  year    = {2012},
  volume  = {12},
  number  = {7},
  pages   = {1119--1141},
  doi     = {10.1080/14697688.2010.494613}
}

\clearpage
\appendix

\section{Supplementary numerical analyses}\label{app:numerical}

\subsection{Window sensitivity}\label{app:calib_audit}

Table~\ref{tab:public_sens} summarizes the window sensitivity. With 126-day windows, the stress target $\rho^G_{\mathrm{target}}$ ranges from 0.459 to 0.550 and $\eps_{\mathrm{calib}}$ ranges from 0.0035 to 0.0048. With 252-day windows, the target is 0.472--0.482. With 756-day windows, the target range falls to about 0.362--0.368 and the radius range to about 0.0021--0.0022. The column $n_{\mathrm{stress}}$ is the number of windows classified as high-stress.

\begin{table}[H]
\centering
\caption{Public-data window sensitivity across rolling-window rules.}
\label{tab:public_sens}
\begin{tabular}{lllll}
\toprule
Window (trading days) & $\rho^G_{\mathrm{target}}$ range & KL radius range & KL increment range  & $n_{\mathrm{stress}}$ range \\
\midrule
126 & 0.459--0.550 & 0.0035--0.0048 & 2.02e-05--2.37e-05  & 13--25 \\
252 & 0.472--0.482 & 0.0037--0.0038 & 2.07e-05--2.11e-05  & 12--26 \\
756 & 0.362--0.368 & 0.0021--0.0022 & 1.57e-05--1.60e-05  & 11--31 \\
\bottomrule
\end{tabular}

\end{table}

\subsection{Benchmark-stress-level sensitivity}\label{app:stress_level_sensitivity}

Table~\ref{tab:stress_level_sensitivity} repeats the policy comparison for benchmark-stress levels $\rho^G_0=0.2,0.4,0.6$ under $\lambda_{\mathrm{norm}}=1$. The selected policy region is stable in the high- and medium-liquidity environments. In the low-liquidity environment, the selected band remains wide; the highest benchmark-stress level selects $b=0.2$ instead of $b=0.3$, consistent with the flat wide-band objective surface.

\begin{table}[H]
\centering
\caption{Sensitivity to the benchmark-stress level under the fixed benchmark-stress view, $\lambda_{\mathrm{norm}}=1$.}
\label{tab:stress_level_sensitivity}
\begin{tabular}{lrrrrrr}
\toprule
Liquidity environment & $\rho^G_0$ & $b^\star$ & $\overline{\HVA}_{\varepsilon_{\mathrm{req}}}(b^\star)$ & $\Delta_b^{\KL}$ & $\Delta_b^{\KL}/\overline{\HVA}_0$ (\%) & $\TE(b^\star)$ \\
\midrule
High-liquidity & 0.2 & 0.02 & \num{0.00062} & \num{1.26e-05} & 2.08 & \num{0.0085} \\
High-liquidity & 0.4 & 0.02 & \num{0.00063} & \num{2.28e-05} & 3.77 & \num{0.0085} \\
High-liquidity & 0.6 & 0.02 & \num{0.00065} & \num{3.23e-05} & 5.33 & \num{0.0085} \\
\midrule
Medium-liquidity & 0.2 & 0.1 & \num{0.0078} & \num{0.00049} & 6.63 & \num{0.065} \\
Medium-liquidity & 0.4 & 0.1 & \num{0.0083} & \num{0.00093} & 12.76 & \num{0.065} \\
Medium-liquidity & 0.6 & 0.1 & \num{0.0087} & \num{0.00137} & 18.72 & \num{0.065} \\
\midrule
Low-liquidity & 0.2 & 0.3 & \num{0.1282} & \num{0.01339} & 11.66 & \num{0.1547} \\
Low-liquidity & 0.4 & 0.3 & \num{0.1396} & \num{0.02478} & 21.58 & \num{0.1547} \\
Low-liquidity & 0.6 & 0.2 & \num{0.1587} & \num{0.03456} & 27.85 & \num{0.1505} \\
\bottomrule
\end{tabular}

\end{table}

\subsection{Worst-case weights and joint-tail amplification}\label{app:tail_measures}

This subsection reports weight-based measures for the KL tilt. In the fixed-radius view, let $w_j^\star(b,\eps)$ denote the worst-case weights from Eq.~\eqref{eq:weights}. The effective sample size is given by
\begin{equation}\label{eq:ess}
  \ESS(b,\eps)=\frac{1}{\sum_{j=1}^N(w_j^\star(b,\eps))^2}.
\end{equation}
This represents the inverse concentration index of the worst-case weights. It equals $N$ when all paths have equal weight and becomes smaller when the robust HVA depends on fewer heavily weighted paths.

Let $D_j(b)$ denote the discounted total turnover on path $j$, and let $M_j$ be the corresponding turnover-weighted average illiquidity multiplier. Define the baseline $(95\%,95\%)$ joint-tail event as
\begin{equation}\label{eq:tail_event}
  A_{95,95}(b)
  =
  \{D_j(b)\ge q_{0.95}^{0}(D(b))\}
  \cap
  \{M_j\ge q_{0.95}^{0}(M)\},
\end{equation}
where the quantiles are computed under the baseline empirical weights. The joint-tail amplification ratio is defined as
\begin{equation}\label{eq:tail_ratio}
  R_{95,95}(b,\eps)
  =
  \frac{\sum_j w_j^\star(b,\eps)\mathbf 1\{(D_j(b),M_j)\in A_{95,95}(b)\}}
       {N^{-1}\sum_j \mathbf 1\{(D_j(b),M_j)\in A_{95,95}(b)\}}.
\end{equation}
Here, the numerator gives the probability of the same baseline-defined joint-tail event under the worst-case weights $Q^\star$, while the denominator gives its baseline empirical probability under $P$. Values above one indicate that the worst-case weights assign more mass to paths where turnover and illiquidity are jointly high. Confidence intervals in Table~\ref{tab:jointtail_ci} are obtained by resampling paths and renormalizing the worst-case weights within each resample; the implementation uses 200 resamples and 95\% intervals.

\begin{table}[H]
\centering
\caption{Joint-tail amplification $Q^\star(A)/P(A)$ for the baseline $(95\%,95\%)$ turnover--illiquidity event.}
\label{tab:jointtail_ci}
{\small
\begin{tabular}{lrp{0.30\linewidth}p{0.30\linewidth}}
\toprule
Liquidity environment & Band & Spread-cost $Q^\star(A)/P(A)$ (95\% CI) & Impact-cost $Q^\star(A)/P(A)$ (95\% CI) \\
\midrule
High-liquidity & 0.00 & 1.363 [1.341, 1.388] & 1.148 [1.120, 1.179] \\
High-liquidity & 0.02 & 1.377 [1.352, 1.399] & 1.162 [1.132, 1.192] \\
Medium-liquidity & 0.00 & 1.320 [1.306, 1.336] & 1.397 [1.379, 1.415] \\
Medium-liquidity & 0.10 & 1.336 [1.310, 1.361] & 1.438 [1.418, 1.459] \\
Low-liquidity & 0.00 & 1.204 [1.191, 1.216] & 1.353 [1.333, 1.373] \\
Low-liquidity & 0.30 & 1.349 [1.295, 1.419] & 1.395 [1.369, 1.427] \\
\bottomrule
\end{tabular}
}

\end{table}

\subsection{Tracking-error-weight sensitivity}\label{app:lambda_sensitivity}
\begin{figure}[H]
\centering
\includegraphics[width=0.86\linewidth]{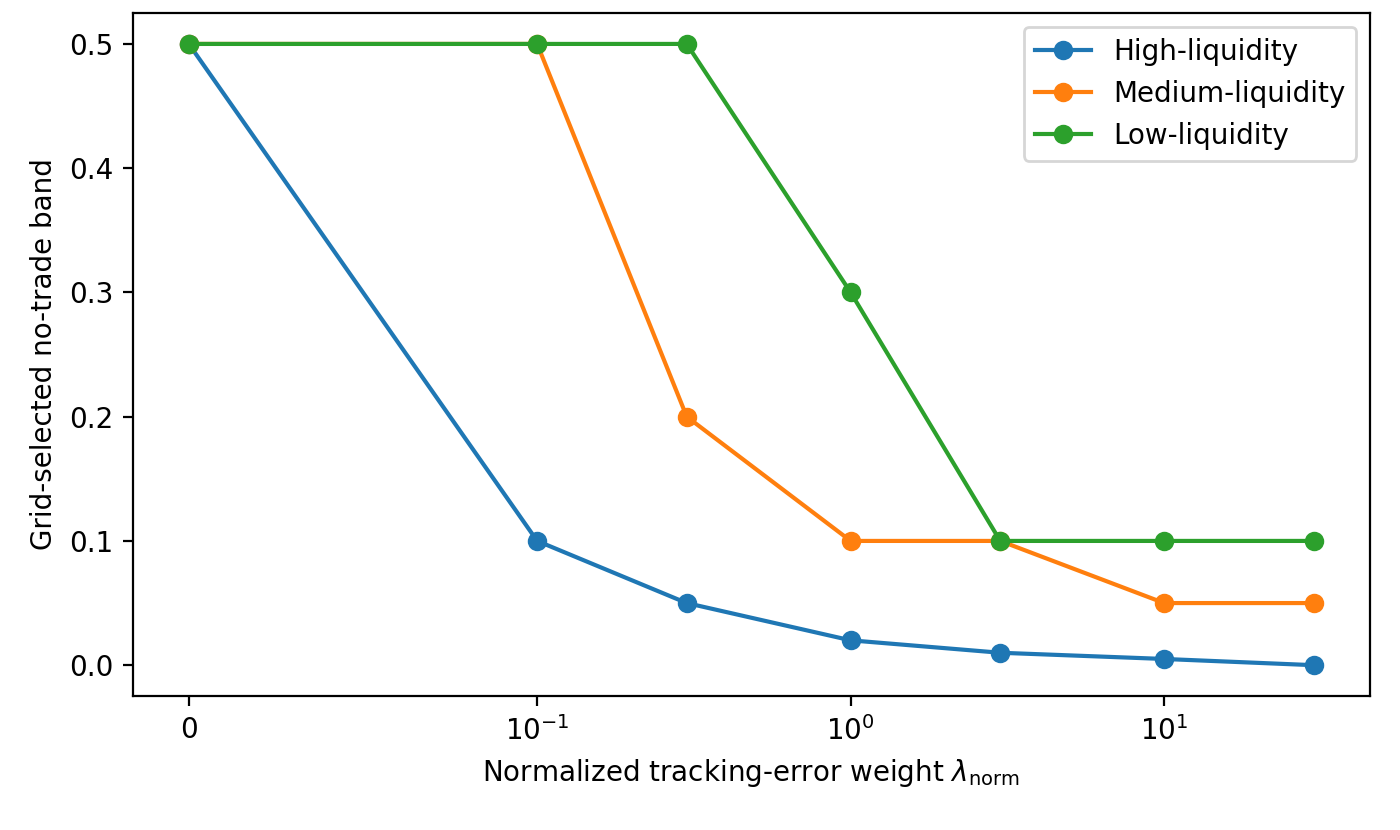}
\caption{Grid-selected no-trade band as a function of $\lambda_{\mathrm{norm}}$ under $\rho^G_0=0.4$.}
\label{fig:lambda_sens}
\end{figure}

Table~\ref{tab:lambda_sens} and Figure~\ref{fig:lambda_sens} present an additional sensitivity analysis. As $\lambda_{\mathrm{norm}}$ increases, the preferred band becomes narrower since tracking error carries greater weight. The low-liquidity environment has wider bands across a broad range of $\lambda_{\mathrm{norm}}$.

\begin{table}[H]
\centering
\caption{Grid-selected band under the fixed benchmark-stress view for alternative tracking-error weights.}
\label{tab:lambda_sens}
\resizebox{\textwidth}{!}{{\small
\begin{tabular}{lrrrrrrr}
\toprule
Liquidity environment & $\lambda_{\mathrm{norm}}=0$ & $\lambda_{\mathrm{norm}}=0.1$ & $\lambda_{\mathrm{norm}}=0.3$ & $\lambda_{\mathrm{norm}}=1$ & $\lambda_{\mathrm{norm}}=3$ & $\lambda_{\mathrm{norm}}=10$ & $\lambda_{\mathrm{norm}}=30$ \\
\midrule
High-liquidity & 0.5 & 0.1 & 0.05 & 0.02 & 0.01 & 0.005 & 0 \\
Medium-liquidity & 0.5 & 0.5 & 0.2 & 0.1 & 0.1 & 0.05 & 0.05 \\
Low-liquidity & 0.5 & 0.5 & 0.5 & 0.3 & 0.1 & 0.1 & 0.1 \\
\bottomrule
\end{tabular}
}
}
\end{table}

\section{Gaussian benchmark construction}\label{app:gaussian_benchmark}

This appendix defines the Gaussian benchmark curve used to translate a KL stress increment into the benchmark-stress label \(\rho^G\). The Gaussian copula is used only as a rank-coupling device: it provides a reproducible benchmark increment curve against which the KL stress increment can be reported. The parameter \(\rho^G\) below is the Gaussian rank-correlation parameter used to construct that curve. Appendix~\ref{app:algorithms} gives the computational steps for the fixed-radius and fixed benchmark-stress calculations.

For path $j$ and band $b$, define the discounted linear and quadratic turnover summaries
\begin{align}
  D^{(1)}_{j,b} &= \sum_{i=1}^{n} e^{-r t_i} X_{j,i}(b),
  &
  D^{(2)}_{j,b} &= \sum_{i=1}^{n} e^{-r t_i} X_{j,i}(b)^2 .
\end{align}
Define the corresponding effective illiquidity multipliers by
\begin{align}
  M^{(1)}_{j,b}
  &=
  \frac{\sum_{i=1}^{n} e^{-r t_i} m_{j,i} X_{j,i}(b)}
       {\sum_{i=1}^{n} e^{-r t_i} X_{j,i}(b)},
  &
  M^{(2)}_{j,b}
  &=
  \frac{\sum_{i=1}^{n} e^{-r t_i} m_{j,i} X_{j,i}(b)^2}
       {\sum_{i=1}^{n} e^{-r t_i} X_{j,i}(b)^2},
\end{align}
with the multiplier set to one when the denominator is zero.
This convention does not affect the cost when there is no turnover. These definitions rewrite the path loss as
\begin{equation}\label{eq:path_cost_decomp}
  L_j(b)=sD^{(1)}_{j,b}M^{(1)}_{j,b}
       +\kappa D^{(2)}_{j,b}M^{(2)}_{j,b}.
\end{equation}

For a candidate benchmark-stress level $\rho^G\in[0,0.99]$, draw $(Z_{1j},Z_{2j})$, $j=1,\ldots,N_G$, from a centered bivariate normal distribution with unit variances and correlation $\rho^G$, and set $U_{kj}=\Phi(Z_{kj})$. The benchmark variables are empirical-quantile transforms,
\begin{align}
  D^{(\ell),G}_{j,b}(\rho^G)&=Q_{D^{(\ell)}_b}(U_{1j}),
  &
  M^{(\ell),G}_{j,b}(\rho^G)&=Q_{M^{(\ell)}_b}(U_{2j}),
  \qquad \ell\in\{1,2\},
\end{align}
where $Q_X$ denotes the empirical quantile function of the path-level sample $X$. The same demand rank $U_{1j}$ is used for $D^{(1)}$ and $D^{(2)}$, and the same liquidity rank $U_{2j}$ is used for $M^{(1)}$ and $M^{(2)}$. Thus, the benchmark modifies the demand–liquidity pairing while keeping the observed marginal scales of the turnover and liquidity.

Then the benchmark cost for path $j$ is
\begin{equation}\label{eq:benchmark_cost}
  C^G_{j,b}(\rho^G)
  =sD^{(1),G}_{j,b}(\rho^G)M^{(1),G}_{j,b}(\rho^G)
   +\kappa D^{(2),G}_{j,b}(\rho^G)M^{(2),G}_{j,b}(\rho^G)
\end{equation}
and the Gaussian benchmark increment is
\begin{equation}\label{eq:benchmark_increment}
  \Delta_b^{G}(\rho^G)
  =
  \frac{1}{N_G}\sum_{j=1}^{N_G} C^G_{j,b}(\rho^G)
  -
  \frac{1}{N_G}\sum_{j=1}^{N_G} C^G_{j,b}(0).
\end{equation}
The map $\rho^G\mapsto\Delta_b^{G}(\rho^G)$ may not be perfectly monotone, so the inverse map uses the monotone envelope
\begin{equation}\label{eq:benchmark_envelope}
  \widetilde\Delta_b^{G}(\rho^G_j)
  =\max_{k\le j}\Delta_b^{G}(\rho^G_k)
\end{equation}
for an ordered grid $\rho^G_1<\cdots<\rho^G_J$. Linear interpolation on this envelope defines the inverse reporting metric. If the target lies within the evaluated grid, interpolation is then applied. The resulting inverse value is denoted $\rho^G_{\mathrm{eq}}$ in the main text.

\section{Computational algorithms}\label{app:algorithms}

\subsection{Loss-sample generation with maturity unwind}

Algorithm~\ref{alg:loss} produces the HVA loss sample. The maturity unwind at the final step is treated separately from the mid-period band updates.

\begin{algorithm}[H]
\caption{Loss samples under no-trade-band delta hedging}\label{alg:loss}
\begin{algorithmic}[1]
\Require Band $b$, time grid $t_0,\ldots,t_n$, price path $S_i$, target delta $\widehat\Delta_i$, discount factors $B_i=e^{-rt_i}$, cost parameters $(s,\kappa)$, illiquidity multipliers $m_i$
\Ensure Path loss $L$, discounted turnover summaries
\State $\phi_0\gets \widehat\Delta_0$, $L\gets0$, $\mathrm{turn1}\gets0$, $\mathrm{turn2}\gets0$
\For{$i=1,\ldots,n-1$}
  \If{$|\widehat\Delta_i-\phi_{i-1}|>b$}
    \State $\phi_i\gets \widehat\Delta_i$
  \Else
    \State $\phi_i\gets \phi_{i-1}$
  \EndIf
  \State $X_i\gets S_i|\phi_i-\phi_{i-1}|$
  \State $L\gets L+B_i\,m_i(sX_i+\kappa X_i^2)$
  \State $\mathrm{turn1}\gets \mathrm{turn1}+B_iX_i$, \quad $\mathrm{turn2}\gets \mathrm{turn2}+B_iX_i^2$
\EndFor
\State \textbf{Maturity unwind:} $\phi_n\gets0$, $X_n\gets S_n|\phi_n-\phi_{n-1}|$
\State $L\gets L+B_n\,m_n(sX_n+\kappa X_n^2)$
\State $\mathrm{turn1}\gets \mathrm{turn1}+B_nX_n$, \quad $\mathrm{turn2}\gets \mathrm{turn2}+B_nX_n^2$
\end{algorithmic}
\end{algorithm}

\subsection{KL upper bound and weights}

\begin{algorithm}[H]
\caption{KL-robust upper HVA and worst-case weights}\label{alg:kl}
\begin{algorithmic}[1]
\Require Loss samples $L_1,\ldots,L_N$, radius $\eps\ge0$
\Ensure Robust upper expectation, realized KL, ESS, weights
\If{$\eps=0$}
  \State $w_j\gets 1/N$ for all $j$
  \State $\overline{\HVA}_0\gets N^{-1}\sum_jL_j$, \quad $\ESS\gets N$, \quad realized KL $\gets0$
\Else
  \State Minimize $g(\theta)=\theta\eps+\theta\log(N^{-1}\sum_j\exp(L_j/\theta))$ over $\theta>0$
  \State Set $w_j\propto\exp(L_j/\theta^\star)$ using log-sum-exp normalization
  \State $\overline{\HVA}_\eps\gets\sum_jw_jL_j$
  \State realized KL $\gets\sum_jw_j\log(Nw_j)$
  \State $\ESS\gets(\sum_jw_j^2)^{-1}$
\EndIf
\end{algorithmic}
\end{algorithm}

\subsection{Fixed-radius and benchmark-stress map}

This algorithm applies the Gaussian benchmark curve from Appendix~\ref{app:gaussian_benchmark} and computes both the fixed-radius and fixed benchmark-stress quantities.

\begin{algorithm}[H]
\caption{Fixed-radius and fixed benchmark-stress map}\label{alg:stressmap}
\begin{algorithmic}[1]
\Require Band grid $\mathcal B$, KL-radius grid $\mathcal E$, fixed KL radius $\eps_{\mathrm{fixed}}$, target stress $\rho^G_0$, loss samples $L(b)$, path-level summaries $(D^{(1)},D^{(2)},M^{(1)},M^{(2)})$
\For{$b\in\mathcal B$}
  \State Compute $\Delta_b^{\KL}(\eps)$ for all $\eps\in\mathcal E$ using Algorithm~\ref{alg:kl}
  \State Compute benchmark increments $\Delta_b^{G}(\rho^G)$ over a grid of benchmark-stress levels $\rho^G$
  \State Form the monotone envelope $\widetilde\Delta_b^G(\rho^G)$ from Eq.~\eqref{eq:benchmark_envelope}
  \State Evaluate the fixed-radius view: compute $\Delta_b^{\KL}(\eps_{\mathrm{fixed}})$ and interpolate $\rho^G_{\mathrm{eq}}(b;\eps_{\mathrm{fixed}})$ from Eq.~\eqref{eq:stress_equiv}
  \State Evaluate the fixed benchmark-stress view: interpolate $\eps_{\mathrm{req}}(b;\rho^G_0)$ from the curve $\eps\mapsto\Delta_b^{\KL}(\eps)$ using Eq.~\eqref{eq:eps_req}
  \State If the target lies outside the evaluated grid, report a boundary value
\EndFor
\end{algorithmic}
\end{algorithm}

\end{document}